# Kernel-level Rootkit Detection, Prevention and Behavior Profiling: A Taxonomy and Survey

Mohammad Nadim

Department of Electrical and Computer Engineering, University of Texas at San Antonio, San Antonio, Texas, USA. Email: mohammad.nadim@my.utsa.edu

Wonjun Lee

The Katz School of Science and Health, Yeshiva University, New York City, New York, USA.

David Akopian

Department of Electrical and Computer Engineering, University of Texas at San Antonio, San Antonio, Texas, USA.

One of the most elusive types of malware in recent times that pose significant challenges in the computer security system is the kernel-level rootkits. The kernel-level rootkits can hide its presence and malicious activities by modifying the kernel control flow, by hooking in the kernel space, or by manipulating the kernel objects. As kernel-level rootkits change the kernel, it is difficult for user-level security tools to detect the kernel-level rootkits. In the past few years, many approaches have been proposed to detect kernel-level rootkits. It is not much difficult for an attacker to evade the signature-based kernel-level rootkit detection system by slightly modifying the existing signature. To detect the evolving kernel-level rootkits, researchers have proposed and experimented with many detection systems. In this paper, we survey traditional kernel-level rootkit detection mechanisms in literature and propose a structured kernel-level rootkit detection taxonomy. We have discussed the strength and weaknesses or challenges of each detection approach. The prevention techniques and profiling kernel-level rootkit behavior affiliated literatures are also included in this survey. The paper ends with future research directions for kernel-level rootkit detection.

## 1 INTRODUCTION

The kernel is a core part of the computer operating system (OS) that plays an important role in managing computer resources. To conduct high privileged arbitrary malicious operations, attackers compromise the OS kernel by loading a malicious kernel module (kernel-level rootkit) into the kernel space. The kernel-level rootkits are the most sophisticated and destructive tools for attackers, because of its nature to hide its presence and obtained high or root privilege. Generally, it is difficult for an ordinary user to find the presence of the kernel rootkit in the system. The lack of protection and isolation in kernel space makes it vulnerable against kernel-level rootkit attack that can perform many malicious operations such as, process hiding, module hiding, network communication hiding, sensitive information gathering, and so on. Because the kernel is the lowest level of an operating system and has highest privileges to access resources, the attacker can access the resources of an operating system by exploiting kernel vulnerabilities. Recently, the kernel-level rootkit technique is employed by more and more malware to gain high privilege in the OS kernel so that they can hide their malicious activities. ZeroAccess malware used rootkit techniques to hide itself in an infected machine and was used to download other malware form a botnet [1]. It infected over millions of Microsoft windows operating system machines. Zacinlo malware leverages rootkit technique to propagate adware in Windows 10 operating system [2].

The detection module of kernel-level rootkit can be located at different layers of a system. Based on the location of the detection module, the mechanisms for the kernel-level rootkit detection can be grouped into three categories: Host-based, Virtualization-based, and External hardware-based. Starting from primitive host-based detection method, virtualization-based detection mechanisms have gained popularity replacing the host-based mechanisms because host-based methods are vulnerable to the kernel-level rootkit. Though hardware-based detection techniques show a good performance, they require expense of great cost. The detection method of kernel-level rootkit can be temporally classified into two different categories: static method and dynamic method. The static method classifies the malicious kernel drivers or modules by analyzing the code to distinguish malicious behavioral features. However, in some cases, obfuscation of code makes it difficult to statically analyze the kernel module, thus dynamic methods are proposed to address the obfuscation problem. The basic idea of detecting kernel-level rootkit by using the dynamic method is to execute the kernel-level rootkit in a proper environment and observe the run-time behavior. The observed run-time behavior is used as a signature to detect kernel-level rootkit in production environment. Some existing techniques use an emulator to execute kernel-level rootkit with some limitations in which, the kernel-level rootkits may not behave correctly in the emulator if they rely on the specific hardware devices. Another approach for the kernel-level rootkit execution is to create virtual machines with full operating system capabilities. Based on working principle, the kernel-level detection approaches can also be classified as signature-based, behavior-based, cross-view based, and integrity-based. A kernel-level rootkit can be detected by monitoring the kernel data structure invariants and creating hypothesized signatures. Hardware events occurred during the execution of system calls in a legitimate and infected system show the behavior of a kernel-level rootkit. The fingerprints of kernel-level rootkit infection can also be traced from the volatile memory to make a cross-view detection. Access control policy can be implemented to enforce the integrity protection of OS kernel against the kernel-level rootkit. The researchers are also focusing on learning-based detection techniques to detect kernel-level rootkit because machine learning and deep learning technology have proven high accuracy to automatically detect known and unknown malware.

Several works have been introduced to survey the prior malware analysis, classification, and detection techniques [3, 4, 5]. According to the interaction with operating system, Rutkowska [13] proposed a classification taxonomy of stealthy malware. Though kernel-level rootkit is a part of the malware family, it is highly distinct from other types of malware. Advantages and disadvantages of technologies to write and detect kernel-level rootkits are briefly discussed in [6]. Tyler Shields [7] presented a brief history as well as the evolution of the rootkits overviewing the detection techniques of different types of kernel-level rootkits including application-level, library-level, firmware-level, and virtualized rootkits. Finally, in the Shields' paper [7], the impact on the digital forensics process that rootkits have was analyzed. A comprehensive and structured view of the prior kernel-level rootkit detection mechanisms was documented by Joy et al [8]. The authors classified the detection mechanism into three different categories based on the position of detection module. A survey on rootkit techniques is detailed by Kim et al [9]. In this survey, both user-level and kernel-level rootkit techniques are described utilizing rootkit samples and different hooking techniques like SSDT hooking, IDT hooking, Inline function hooking are briefly described by the authors. Bravo and Garcia [10] discussed the classification and techniques of rootkit followed by the rootkit detection approaches. Li et al. [11] surveyed the core implementation details of kernel malware by studying several Linux kernel malwares. Rudd et al [12] surveyed the stealth technologies highly adopted by the kernel-level rootkits with detailed discussion. They discussed different types of hooking techniques as well as the DKOM technique. Not only the stealth techniques but also their countermeasures are overviewed in this paper. Most importantly, prior machine learning-based countermeasures to detect stealth techniques are discussed briefly. The authors also identified some flawed algorithmic assumptions that hinder malware recognition in the machine learning approach.

## 1.1 Problem Statement

Though the kernel-level rootkit attack number is small compared to all reported malware infections, the impact of the kernel-level rootkit is fairly large in terms of malicious activities. The elusive nature of kernel-level rootkit makes it difficult to detect, still different approaches have been introduced to detect kernel-level rootkit. There has been a lack of work that details most of the contemporary research affiliated to the kernel-level rootkit detection techniques in a structured way. Also, a comparison of strength and weakness / challenge between different detection approaches need to be addressed. The state-of-the-art research on the kernel-level rootkit prevention along with behavior profiling are required to be discussed in detail.

## 1.2 Contribution

The contribution of this study briefly is:

1. This survey is an endeavor to provide a broad and structured overview of extensive research on the kernel-level rootkit detection techniques.
2. We have proposed a solution taxonomy on the kernel-level rootkit detection mechanism (figure 1).
3. Strength and weakness are compared between different kernel-level rootkit detection approaches.
4. Learning-based techniques for kernel-level rootkit detection are widely detailed in this study.
5. Profiling the elusive nature of kernel-level rootkit behavior affiliated prior literatures are included in this survey along with the contemporary research on kernel-level rootkit prevention techniques.

The rest of the paper is organized as follows: Section 2 briefly describes the kernel-level rootkit attack approaches; Section 3 categorizes kernel-level rootkit detection techniques in the literature. An overview of the kernel-level rootkit prevention techniques, existing literatures to profile kernel-level rootkit behavior are described in Section 4. Future research directions are described in Section 5 and Section 6 concludes this survey paper.

## 2 KERNEL-LEVEL ROOTKIT

The first generation of rootkits are mainly user-level rootkits that conceal themselves as disk-resident system programs by mimicking the system process files. Those rootkits are easy to detect and remove by using file integrity tools and user-level security software. So, the modern rootkits have evolved from disk-residency to memory-residency to evade the detection by file integrity tools. The second generation of rootkits modify the control flow of the computer system to execute malicious code by using different hooking techniques. The return value or functionality requested from the operating system can be altered by executing the malicious code. User-mode hooking is comparatively easier to detect than kernel-mode hooking, as it is implemented in the user-space. Kernel-mode hooking usually injects malicious code into the kernel-space of an OS via device driver which makes it difficult to detect by user-mode intrusion detection system (IDS) and other security tools. System Service Descriptor Table (SSDT), Interrupt Descriptor Table (IDT) and I/O Request Packet (IRP) function tables are the most common target for implementing kernel hooks. The execution of malicious code by the second-generation rootkit leaves memory footprint in both user-space and kernel-space that can be detected and analyzed. The third generation of rootkits are mostly kernel-level rootkits. In spite of having limited applications, but they are difficult to detect as they can modify the dynamic kernel data structures. Direct Kernel Object Manipulation (DKOM) attack, implemented by the third-generation rootkits, targets the dynamic data structures in kernel whose values change during runtime. Kernel-level rootkit can be summarized into the following categories: System Service Hijacking (system call table hooking, replacing system call table), Dynamic Kernel Object Hooking (virtual file system hooking), and Direct Kernel Object Manipulation (DKOM).

## 2.1 System Service Hijacking

A system call is basically an interface between user level processes and an operating system. User level programs access the system resources through this interface. All the actual system call routine addresses are stored in a table called system call table or system service descriptor table. The system calls can be differently attacked by the kernel-level rootkits. For example, attackers can replace the legitimate system call with own malicious system call by modifying the system call address in system call table. Attackers can also change the control flow of a system call by modifying the code in the target address. Usually by inserting jump instructions, the control is passed to the malicious code. Additionally, the whole system call table can be replaced by attackers with own version of system call table by overwriting the memory that contains the system call table address [19]. Another important hooking target is the Interrupt Descriptor Table (IDT). The processor uses the IDT to determine the correct response to interrupts and exceptions. As interrupts have no return values, interrupt requests can only be denied by hooking the IDT. In a multiprocessing system, an attacker needs to hook all IDTs as each CPU has its own IDT.

## 2.2 Dynamic Kernel Object Hooking

The OS kernel uses Virtual File System (VFS) to handle the file system operations across different types of file systems such as EXT2, EXT3, and NTFS. Thus, VFS is a layer between the actual file systems and the user-level programs that make the file handling system calls to access the files. Different data structures are used by VFS to achieve a common file model such as the file object, inode object, and dentry object. The kernel-level rootkit can modify the file object data structure field that contains a pointer to the file_operation structure (f_op) to hide without modifying the system call table. Function pointers to inode operation functions such as lookup function are stored in the inode data structure. The kernel-level rootkit can hide a process by modifying the function pointer of the lookup function for the process directory's (/proc) inode data structure [14].

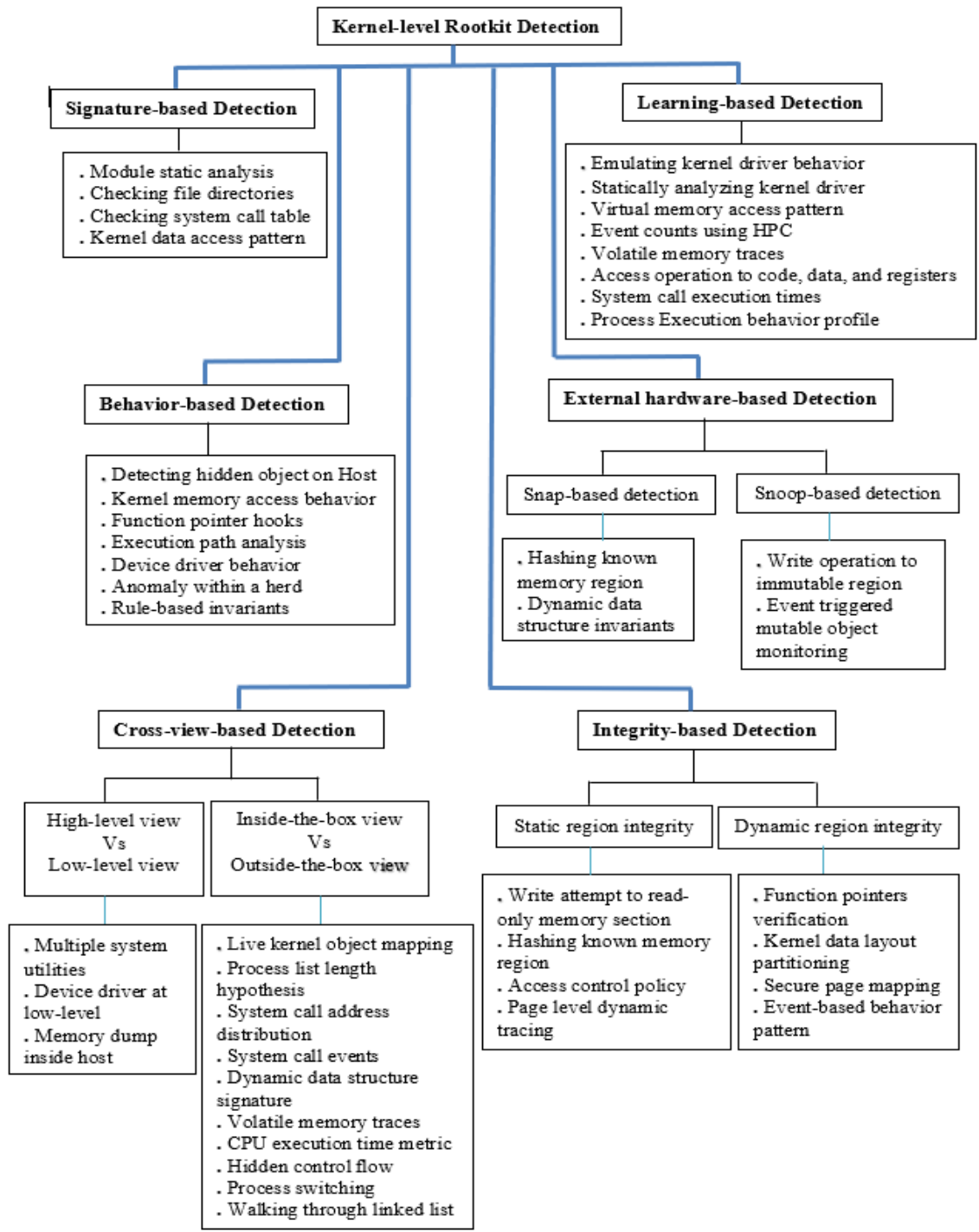

Figure 1: Proposed taxonomy of the Kernel-level rootkit detection approaches.

## 2.3 Direct Kernel Object Manipulation (DKOM)

Kernel-level rootkits can also modify the kernel data structure by using DKOM technique. As DKOM technique aims to modify dynamic kernel data structures, it is harder to detect than kernel hooking because the dynamic object changes during normal runtime operations. Malicious process hiding is a perfect example of DKOM technique. In Windows OS, an _EPROCESS data structure is associated with each process. To hide a malicious process, kernel-level rootkits modify the _EPROCESS data structure that is maintained in a doubly linked list. Unlinking an element from the process list implemented in _EPROCESS data structure makes the process invisible to both user and kernel mode programs. Other than process to hide itself with the DKOM techniques, Kernel device drivers, active ports can also be hidden by using this technique. Implementation of DKOM is extremely difficult because incorrect change in operating system kernel data structure may result in system crashes.

Table 1 summarizes the kernel-level rootkit detection approaches selected for this study based on environment (Host, Virtual Machine, Emulator), focused feature (Static, Dynamic) and operating system (Windows, Linux, macOS).

Table 1: Summary of the Kernel-level rootkit detection approaches selected for this study.

| Detection Approach | Prior Works | Environment | | | Focused Feature | | Operating System | | |
|---|---|---|---|---|---|---|---|---|---|
| | | Host | Virtual Machine | Emulator | Static | Dynamic | Windows | Linux | MacOS |
| Signature-based | Kruegel et al. [15], Levine et al. [19, 20, 21], KRGuard [23, 24] | √ | | | √ | | | √ | |
| | Zhou and Makris [22] | | | √ | √ | | | √ | |
| | DataGene [25, 26] | | | √ | | √ | | √ | |
| Behavior-based | Ring and Cole [27], DCFI-Checker [35] | √ | | | | √ | | √ | |
| | KernelGuard [28] | | √ | | √ | √ | | √ | |
| | HookScout [29] | | √ | | √ | √ | √ | | |
| | Numchecker [31, 32], Wang et al. [33], KLrtD [41] | | √ | | | √ | | √ | |
| | Patchfinder [34] | √ | | | √ | | | √ | |
| | Blacksheep [37], dAnubis [36] | | | √ | | √ | √ | | |
| | Fluorescene [38] | | √ | | | √ | √ | √ | |
| | Wang [39] | √ | | | | √ | √ | | |
| Cross-view-based | Strider GhostBuster [45] | √ | √ | | | √ | √ | √ | |
| | Wampler and Graham [50, 51] | √ | | | √ | | | √ | |
| | Molina et al. [42], KeRTD [43], Rkfinder [60], HyBIS [63], WinWizard [64], Dolan-Gavitt et al. [55] | | √ | | | √ | √ | | |
| | Lycosid [49] | | √ | | | √ | √ | √ | |
| | XView [53], SigGENE [56] | | | √ | | √ | √ | | |
| | BeCFI [71] | √ | | | | √ | | √ | |
| | SigGraph [57] | | | √ | √ | √ | | √ | |

Table 1: Continued.

| Category | Method | | | | | | | |
|---|---|---|---|---|---|---|---|---|
| Cross-view-based | DeepScanner [44] | √ | √ | | | √ | | √ | |
| | Xie and Wang [58], Hua and Zhang [62], VMDetector [54], RMVP [72] | | √ | | | √ | | √ | |
| | MASHKA [46] | | √ | | | √ | √ | √ | √ |
| | HyperLink [61] | | | √ | | √ | √ | √ | √ |
| | MAS [66], Zaki and Humphrey [65] | | √ | | √ | √ | √ | | |
| | XenKIMONO [73] | | √ | | √ | √ | | √ | |
| | Case and Richard [67] | | √ | | | √ | | | √ |
| | AUTOTAP [70], LiveDM [47] | | | √ | | √ | | √ | |
| Integrity-based | Pioneer [80], EPA-RIMM [88], SGX-Mon [89] | √ | | | √ | | | √ | |
| | SBCFI [30], Sentry [94, 95], OSck [78], StackSafe [77], Zhan et al. [93], BehaviorKI [97] | | √ | | √ | √ | | √ | |
| | Xu et al. [92], Psyco-Virt [82], Paladin [75, 76], Win et al. [87], CloudMon [91], Livewire [74] | | √ | | √ | | | √ | |
| | Patagonix [84] | | √ | | √ | | √ | √ | |
| | KOP [48] | | √ | | √ | √ | √ | | |
| | Kvm-SMA [86], Zhang et al. [79], RootkitDet [83] | | | √ | √ | √ | | √ | |
| | MOSKG [96] | | √ | | √ | √ | √ | √ | |
| External hardware-based | Copilot [98], Vigilare [104, 105], GRIM [100] | √ | | | √ | | | √ | |
| | Petroni et al. [101], Wang and Dasgupta [99], Gilbraltar [102, 103], KI-Mon [106, 107] | √ | | | √ | √ | | √ | |
| Learning-based | Limbo [108] | | | √ | √ | √ | √ | | |
| | Musavi and Kharrazi [109] | √ | | | √ | | √ | | |
| | Luckett et al. [114] | | √ | | | √ | | √ | |
| | Xu et al. [110], Zhou and Makris [115] | | | √ | | √ | | √ | |
| | Singh et al. [111], VKRD [113] | | √ | | | √ | √ | | |
| | TKRD [112] | | √ | | | √ | √ | | |

## 3 KERNEL-LEVEL ROOTKIT DETECTION

Kernel-level rootkit detection approaches can be categorized into six major classes: signature-based, behavior-based, cross-view-based, integrity-based, external hardware-based, and learning-based. Then each major category can be sub-categorized according to underlying working principles.

### 3.1 Signature-based Detection

Signature-based detection is one of the most common techniques used to address software threats. This type of detection involves detection tools having a predefined repository of static signatures (fingerprints) that represent known threats. Different signature-based kernel-level rootkit detection techniques are discussed in detail in this section. The strengths and weaknesses or challenges of the signature-based kernel-level rootkit detection approaches are shown in Table 2.

### 3.1.1 Module Static Analysis

The most common way of inserting kernel-level rootkits into the memory is through the loadable kernel module (LKM). The runtime behavior of kernel-level rootkits significantly differs from the one of the regular kernel modules or device drivers. Before loading into the kernel, a module's binary can be checked for malicious instruction sequences signature that either performs write operation to an illegal memory area or calculate an address in the kernel space using a forbidden kernel symbol reference and performs write operation using the calculated address. A similar approach is proposed by Kruegel et al. [15] to detect kernel-level rootkit module by leveraging symbolic execution. This method is ineffective against malicious code injection in the kernel which does not use module loading interface.

### 3.1.2 Checking File Directories

Some primitive detection tools have used to look into file directories for kernel-level rootkit detection since some rootkits create a specific directory name in a certain directory (e.g., 'Knark' rootkit creates a directory named '*/proc/knark*'). Detection is performed by checking some predefined directories. Detection tools like Chkrootkit [16], OSSEC [17] combine file directory signature checking with other techniques to detect kernel-level rootkit. However, this type of detection can be easily evaded by slightly modifying the directory name.

### 3.1.3 Checking System Call Table

As system calls are used to access the system resources, it is the most targeted object by the kernel-level rootkit. System call table data structure stores the system call addresses in the kernel memory. Kernel-level rootkit can tamper system calls in three ways: by modifying the system call address in the system call table to a malicious address; by overwriting first few instructions of the system call with jump instruction to execute malicious code; by redirecting the entire system call table to a new kernel memory location. Samhain Lab developed Kern_check [18] program that can compare current system call table with the original system call table stored in '*/boot/System.map*' system file of Linux OS to detect kernel-level rootkit that overwrite the system call table. Modification of system call is complicated due to rare condition. By comparing with hash values of uninfected system call can indicate a modification. Levine et al. [19] modified kern_check program to detect the system call table redirection. They assumed that the implementation of each malicious system call is unique for particular kernel-level rootkit resulting in signature that can be used to categorize the kernel-level rootkits [20, 21]. Zhou and Makris [22] used several x86 hardware conventions to detect system call table and system call routine modification. KRGuard [23, 24] uses recent hardware feature of the processor to detect kernel-level rootkit that modify the system call table. However, in this technique, it is not possible to detect DKOM attack for its nature not to affect the system calls.

### 3.1.4 Kernel Data Access Pattern

A Kernel-level rootkits have evolved from injecting malicious code to maliciously reusing legitimate code. Unique data patterns exist when kernel-level rootkit tampers with the core kernel data. Kernel memory access information such as accessing code, the accessed memory type, and the accessed offset can create data access behavior signatures. DataGene [25, 26], a data-centric OS kernel malware characterization prototype, analyzes the data access behavior of the dynamic kernel objects of the monitored OS at runtime by using memory allocation events. These data access signatures can be used to detect the classes of kernel-level rootkits that share the same data access pattern. The access patterns are not only common in a similar class of rootkits but also found across a variety of different classes.

Table 2: Signature-based Detection of Kernel-level Rootkit: Strengths and Challenges/Weaknesses.

| Approaches | Strengths | Challenges/Weaknesses |
| --- | --- | --- |
| Module static analysis. | Do not need to load the module. | Increased module loading time. |
| Checking file directories. | Fast detection. | Easy to evade by slight modification. |
| Checking system call table. | Easy to detect modification. | Values need to be stores and DKOM attack cannot be detected. |
| Kernel data access pattern. | Classes of kernel-level rootkit can be detected. | Performance overhead can occur. |

## 3.2 Behavior-based Detection

Behavior-based detection evaluates an attack based on its intended actions or behavior. Attempts to perform actions that are clearly abnormal or unauthorized would indicate the action is malicious, or at least suspicious. Different behavior-based kernel-level rootkit detection techniques are discussed in detail in this section. The strength and weaknesses or challenges of the behavior-based kernel-level rootkit detection approaches are shown in table 3.

### 3.2.1 Detecting Hidden Objects on Host

Intruders often install kernel-level rootkits and later securely remove the binary from the disk to modify the kernel directly in the memory without leaving any trace against the traditional file discovery techniques. This type of rootkits can only be detected by monitoring behaviors of hiding objects like processes, modules, network connections etc. Ring and Cole [27] presented a design of a software-based forensics system that is capable to restore evidence of kernel-level rootkit from volatile memory. The design was implemented as a loadable kernel module to collect all running processes, dynamic kernel memory, system call addresses, all loadable kernel modules, and desired process information. The system freezes the processes, mounts the hard drive in read-only mode, and stores the evidence on a removable media to avoid being corrupted by kernel-level rootkit.

### 3.2.2 Kernel Memory Access Behavior

Static kernel data are easy to determine from the kernel symbol table and can be protected without any sort of tracking by applying policies to any memory writes to the protected memory range. As the dynamic kernel data are dynamically allocated in any portion of the memory, first the location of data needs to be tracked before detecting any illegal memory access. Watchpoints, that watches memory accesses to a pointer to the protected data structure, need to be implemented to track dynamic data structure pointer and the data it points to. Then the illegal memory accesses can be observed by detecting data structure modification from unauthorized function. Based on the characteristics of kernel source code, one can enforce what kernel code is allowed to or prohibited from accessing protected kernel data. KernelGuard [28] is an example of detecting and preventing kernel-level rootkit using kernel memory accesses.

### 3.2.3 Function Pointer Hooks

Kernel-level rootkit can target dynamically allocated function pointers in kernel data structures to modify persistent control flow. The large number of kernel objects and function pointers along with closed-source operating system can make it difficult to generate effective hook detection policy. HookScout [29] used binary code analysis to track function pointers for generating hook detection policy without accessing OS kernel source code.

### 3.2.4 Execution Path Analysis

An analysis [30] on Linux kernel-level rootkits shows that a significant number of kernel-level rootkits persistently violate control-flow integrity. The number of some hardware events occurred during the execution of a kernel function is different if the control-flow of that kernel function is maliciously modified. These events can be easily counted using hardware performance counter (HPC), a part of the performance monitoring unit in most modern processors. NumChecker [31, 32], a virtual machine monitor (VMM) based framework, can detect malicious modification to a system call by control-flow modifying kernel-level rootkits in the guest VM by checking the number of certain hardware events in host OS during system call's execution. Wang et al. [33] extended their hardware performance counter-based kernel-level rootkit detection approach to a new level that locally collect the hardware events sample but remotely analyze it. Remote analyzer reduces the computing resource overhead of the monitored system and compressive sensing technique [137] for compressed fine-grained HPC profiles minimizes the I/O bandwidth required for data transmission. Patchfinder [34], developed by Rutkowski, analyzes the execution path of system calls to calculate the number of instructions used to execute that system call. The number of instructions in an uninfected system needs to be calculated beforehand to compare them with the suspected system. This approach is not suitable to detect DKOM attack. DCFI-Checker [35] checks the dynamic control flow integrity by counting the executed branch instructions using performance monitoring counter.

### 3.2.5 Device Driver Behavior

Kernel-level rootkit typically takes a form of device driver in Windows OS. To detect this type of rootkit, a comprehensive picture of the device driver needs to be provided by observing events such as the execution of driver's code, invocation of kernel functions, and access to the hardware. dAnubis [36] analyzes device driver's behavior by instrumenting the emulation environment and provides a human readable report. Along with common kernel-level rootkit techniques such as hooking, kernel patching and DKOM, dAnubis gives an overview of driver's interaction with other drivers and interface to user-space processes.

### 3.2.6 Anomaly Within a Herd

By taking the advantage of the similarity amongst a group of analogous machines in a distributed system, one can effectively detect anomaly caused by kernel-level rootkit. Physical memory dumps can be used for configuration, kernel code, kernel data and kernel entry points comparison to detect an anomalous machine. As long as the majority of machines are uncompromised and viable memory dumps are available, Blacksheep [37] can distinguish compromised machines and also properly identify anti-virus software, self-modifying code used for security purposes. Fluorescence [38] is a detection approach with limited knowledge of kernel to detect infected virtual machine by kernel-level rootkit within a herd of similar virtual machines. The location of the page global directory and the processor's instruction set are used to concisely fingerprint each kernel. Deep learning and clustering approaches are used in Fluorescence to find out the anomalous virtual machines.

*3.2.7 Rule-based Invariants*

As kernel-level rootkit modifies the kernel data structure and kernel objects, it leaves some inconsistencies in the system. We can define some rules to hold for a clean system and indicate any deviation of the rules as an attack. For an example, we can define a rule such that in Linux OS, *task_struct* and *run_list* both data structures output should be the same. Wang [39] introduced a rule-based approach that chooses different data structures in different layers and performs an information calculation process to define rules as invariants based on the information. KLrtD [41] extracts whitelist rules from normal kernel execution during inference phase and uses those rules for checking data structures integrity violation during integrity checker phase.

Table 3: Behavior-based Detection of Kernel-level Rootkit: Strengths and Challenges/Weaknesses.

| Approaches | Strength | Challenges/Weaknesses |
|---|---|---|
| Detecting hidden objects on host. | Software implementation to store evidence. | Need to rely on host OS. |
| Kernel memory access behavior. | Dynamic data can be protected. | Need OS kernel source code. |
| Function pointers hook. | No need to access OS kernel source code. | Detection system running inside the host can be subverted. |
| Execution path analysis. | Enhanced security with reduced performance overhead. | Vulnerable against DKOM attack. |
| Device driver behavior. | Malicious device driver behavior can be emulated. | Unable to analyze device driver exempt kernel rootkit injection. |
| Anomaly within a herd. | Effective for homogeneous corporate networks and clouds. | Will not work if majority of machines are compromised. |
| Rule-based invariants. | Do not need prior knowledge of kernel-level rootkit. | A large number of invariants set. |

### 3.3 Cross-view-based Detection

The basic idea of cross-view-based detection is to compare two different views of the system. We can divide cross-view-based detection into two sub-categories: high-level view vs. low-level view and inside-the-box view vs. outside-the-box view. In the first category, it is easier to extract the views, but the data can be compromised by the kernel-level rootkit. In the second category, it is difficult to construct the view from outside the box, while the data is safe from the kernel-level rootkit. An overview of cross-view-based detection approach is shown in figure 2. The strength and weaknesses or challenges of the cross-view-based kernel-level rootkit detection approaches are shown in table 4.

*3.3.1 High-level View Vs Low-level View*

*3.3.1.1 Multiple System Utilities*

Any discrepancy between outputs in gathered data by multiple system utilities from user-space could lead to kernel-level rootkit detection. Molina et al. [42] proposed a live forensic tool based on this idea. However, the data of the forensic tool can be compromised by active kernel-level rootkits since the tool is running in user space with a lower privilege than rootkits.

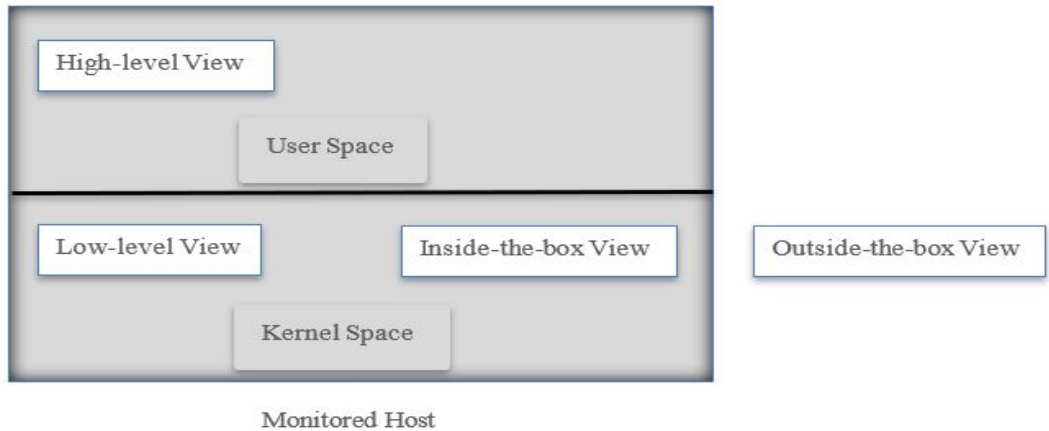

Figure 2: An overview of cross-view-based kernel-level rootkit detection mechanism.

*3.3.1.2  Device Driver at Low-level*

A low-level view of the running system can be portrayed using a device driver implemented in the kernel. But this approach is vulnerable against kernel-level rootkit as both of device driver and rootkit have the same privilege. An access control list can be enforced to avoid the subversion. Kernel Rootkit Trojan Detection (KeRTD) [43], a cross-view-based solution implemented in the host, uses view-difference to detect kernel-level rootkits. DeepScanner [44], implemented as Loadable Kernel Module (LKM) in Linux OS, uses inter-structure signature and imported signature concepts to scan kernel memory for detecting hidden processes, sockets, and kernel modules according to proposed invariants. The output of system utilities including *ps*, *netstat*, and *lsmod* is used for a cross-view comparison to detect kernel-level rootkits. Strider GhostBuster [45] also uses a driver to perform low-level scan and compare the result with a high-level scan.

*3.3.1.3  Memory Dump Inside Host*

Korkin and Nesterov proposed Malware Analysis System for Hidden Knotty Anomalies (MASHKA) [46] for a memory dumping and analysis of a host that can be used to detect kernel-level rootkits. MASHKA uses encryption to protect the saved dump file from modification. The analysis system is implemented in a Windows OS and uses a dynamic bit signature (DBS) to obtain all process lists from dump memory file EPROCESS structure that can be compared with the list obtained by system utility tools. This system is also able to detect hidden drivers. The authors additionally discussed the possibility of MASHKA to be deployed as security as a service (SaaS) in the cloud.

**3.3.2  Inside-the-box View Vs Outside-the-box View**

*3.3.2.1  Live Kernel Object Mapping*

Snapshot-based memory mapping are time specific and kernel memory can be manipulated within the time-gap between two memory snapping by the kernel-level rootkit. And not all the data structures have an invariant to create an untampered view. By capturing the allocation and deallocation events of the kernel object, a live untampered view of that kernel object can be mapped. A difference between the set of kernel object found by traversing the kernel memory and a live untampered view indicates an anomaly caused by kernel-level rootkit. Using this approach, LiveDM [47] detects DKOM-based kernel-level rootkit. KOP [48] has the ability to map the kernel objects that can be used to detect objects hidden by kernel-level rootkit.

*3.3.2.2   Process List Length Hypothesis*

The length of process lists obtained from a low-level and high-level can be compared to detect hidden process by kernel-level rootkit. It is sufficing on an idle system by taking a single instance of the two process lists and compare them. But on an active system, without perfect synchronization there could be false positive results. Lycosid [49] obtains a trusted view of guest processes from within a VMM and overcomes this problem by taking many pairs of measurements over time and then performs a paired sample hypothesis to estimate the number of hidden processes.

*3.3.2.3   System Call Address Distribution*

The knowledge about the distribution of system call addresses in a clean system can be a good measure for detecting kernel-level rootkits. Wampler and Graham [50] proposed a statistical technique that compares the distribution of system call addresses in a clean system and suspicious system. The experiment with a couple of kernel-level rootkits showed that the 'largest extreme value' distribution using Anderson-Darling (AD) test [138] can be used to detect kernel-level rootkit. The authors later experimented with Enyelkm kernel-level rootkit that attacks the system via system call target modification [51]. In system call target modification attack, the system call table does not need to be changed, but only the first few instructions are overwritten with a jump instruction that redirects the control flow to malicious code. The authors first disassembled the running kernel to collect all conditional and unconditional jump instructions and then analyzed the memory address operands of those instructions. The appearance order of these memory addresses is considered as the second dimension. Then a normality-based detection is used to detect the malicious addresses.

*3.3.2.4   System Call Events*

Due to the semantic gap, it is difficult to acquire knowledge about guest kernel data structure from virtual machine monitor and also advanced attacks can tamper the guest kernel data structures layout [52]. Semantic gap problem to reconstruct process information can be overcome by intercepting and interpreting system call events of the guest operating system. Executed instructions can be tracked to intercept the beginning and return of a system call event. Then the parameter along with the system call can be interpreted by reading certain hardware register values. XView [53] constructs an outside-the-box view of active processes list from system call events and compares it with inside-the-box system utility tools output to detect hidden processes. VMDetector [54] uses system call events to construct active processes list from kernel-level view and VMM-level view and then compares it with a user-level view to detect hidden processes.

*3.3.2.5   Dynamic Data Structure Signature*

Kernel-level rootkit often uses a DKOM technique to hide processes, threads, and modules. The hidden objects can be detected by scanning data structure objects signatures in the kernel memory and perform a cross-view detection. Kernel-level rootkit can modify non-essential fields of the data structures to evade the memory scanning detection relying on brittle signatures. The robust signatures of the data structure fields will make the object invalid if changed. A similar work has been proposed by Dolan-Gavitt et al. [55]. The authors have shown that it is possible to evade memory scanning by modifying the non-essential fields of the EPROCESS data structure in Windows OS. The profile of data structure objects' robust fields during execution is also used as signatures to detect kernel-level rootkit. SigGENE [56] profiles kernel object features during malicious code execution. SigGraph [57] generates graph-based structural invariant signatures that can achieve high accuracy in recognizing kernel data structure instances.

*3.3.2.6 Volatile Memory Traces*

Kernel-level rootkit may hide malicious modules, processes, network connections etc., but still it leaves its footprint to volatile memory while it is executed. Kernel-level rootkits that do not use DKOM techniques are easier to detect by simply reconstructing the corresponding data structure's view from volatile memory. For example, *PsActiveProcessHead* and *init_task* are the head of the process list in Windows and Linux OS, respectively. One can go through the complete process list starting from this position. Xie and Wang [58] applied this approach to other data structures to detect kernel-level rootkit. However, this approach is vulnerable against DKOM technique as it modifies the data structures in the memory. Dynamic data structure signatures described in the previous section, 4.3.2.5 are used to locate all data structure objects. Volatility [59] is well-known framework to reconstruct data structure view from volatile memory. Rkfinder [60] generated an abstract view of the system state to reveal the inconsistencies by integrating major capabilities of Volatility framework. The drawback of memory forensic tools is their dependency on up-to-date kernel information of the target OS. HyperLink [61] is an implementation of partial retrieval of process information using memory forensic without requiring OS kernel source code. Other literatures like Hua and Zhang [62], HyBIS [63], WinWizard [64], Zaki and Humphery [65] leverage traces from memory to detect kernel-level rootkit. MAS [66] uses memory traversing to find the visibility of data objects to system tools found from memory snapshot.

Most of the prior research on kernel-level rootkit detection were focused on Windows and Linux-based operating system. Case and Richard [67] proposed new memory forensic and analysis techniques for the Mac OS X system motivated by Windows and Linux-based detection strategies. The authors described the system service functionalities that can be abused and developed Volatility plugin for each of those services to detect tampering or malicious use. Volafox [68] is a memory analysis toolkit for Mac OS X that can be used to detect malicious modification of memory by a kernel-level rootkit. Kyeong-Sik Lee [69], the prime developer of Volafox, described the memory forensic technique adopted by Volafox.

*3.3.2.7 CPU Execution Time Metric*

CPU execution time could be a reliable source for constructing a view of running processes list as it is very critical to forge the value. One can hook the tap points (an execution point where monitoring can be performed) of process data structure object creation and deletion, then count the CPU execution time of the executed process. A hash table can be used to store the accumulated CPU time for each process. AUTOTAP [70] uncovers such tap points for kernel data structure objects. A cross-view comparison between the running process list and the output of system utility can detect hidden process.

*3.3.2.8 Hidden Control Flow*

Kernel-level rootkit introduces unintended or hidden control flow by injecting new instructions or misusing existing instructions. Since every instruction must be issued to the processors, it is impossible for kernel-level rootkit to fool the processor by modifying control flow. One can construct a hardware view of the sum of branch instructions issued to the processors with the support of performance monitoring counter. A cross-view comparison with software view of executed instructions will show the hidden control flow. BeCFI [71] is an implementation of this approach.

*3.3.2.9 Process Switching*

By monitoring the process switching and mapping the memory, it is possible to construct a semantic view of running processes inside a guest VM. One can monitor process switch to check kernel stack switching and extract the corresponding raw memory using memory mapping. Then the raw memory is translated into high-level semantics with the help of a semantic library. RMVP [72] creates a real-time process monitor to detect hidden processes.

*3.3.2.10 Walking Through Linked List*

One can construct a kernel view of loaded modules and a list of running processes by walking through the corresponding linked link. Then the output of system utility tools can be used for a cross-view detection. This approach is vulnerable against DKOM attack, as the kernel-level rootkit unlink the data object from the linked list. XenKIMONO [73] uses this approach for cross-view-based detection along with integrity measurement.

Table 4: Cross-view-based Detection of Kernel-level Rootkit: Strengths and Challenges/Weaknesses.

| Approaches | Strength | Challenges/Weaknesses |
|---|---|---|
| **High-level view Vs Low-level view** | | |
| Multiple system utilities. | Can be implemented inside the host. | Vulnerable against modern kernel-level rootkit. |
| Device driver at low-level. | Scanning can be done in a short time. | Have same privileges as kernel-level rootkit. |
| Memory dump inside host. | Kernel memory can be dumped inside the host with encryption. | Kernel-level rootkit can subvert the detection system. |
| **Inside-the-box view Vs Outside-the-box view** | | |
| Live kernel object mapping. | Untampered view cannot be manipulated by kernel-level rootkit. | Obfuscation technique can confuse the detector. |
| Process list length hypothesis. | Trusted view is constructed outside the host. | Only applicable to detect hidden process. |
| System call address distribution. | Effective for system call target modification attack. | Natural outlier may incur disturbance. |
| System call events. | Active running process list can be monitored. | Unable to detect hidden module. |
| Dynamic data structure signature. | Dynamic kernel objects can be detected. | Signatures can be evaded. |
| Volatile memory traces. | Detection system can be implemented remotely. | Depend on OS kernel information and transient attacks may remain undetected. |
| CPU execution time metric. | Difficult to forge the execution time value. | Need to store a hash table. |
| Hidden control flow. | Impossible to fool the processor by modifying control flow. | Need of hardware support increases overhead. |
| Process switching. | Real time process can be monitored. | Only hidden process can be detected. |
| Walking through linked list. | Kernel objects hidden from system utilities can be easily detected. | Vulnerable against DKOM attack. |

## 3.4 Integrity-based Detection

The kernel-level rootkit tampers the integrity of both static region and dynamic region of the operating system. While some research focuses on only static region integrity, recent research focuses on dynamic region integrity as modern kernel-level rootkits mostly alter the dynamic data structures. It is comparatively easier to check the integrity of static region as the dynamic region changes during runtime operation. The strength and weaknesses or challenges of the integrity-based kernel-level rootkit detection approaches are shown in table 5.

### *3.4.1 Static Region Integrity*

#### *3.4.1.1 Write Attempt to Read-only Memory Section*

In modern computer architecture, certain sections of memory are read-only as a part of memory protection interface. Kernel-level rootkits modify these sections by running with the highest privilege. A significant research in this area was done by Garfinkel and Rosenblum [74]. They built Livewire at hypervisor layer that detects any write attempt to modify the sensitive read-only memory section by leveraging the isolation, inspection, and interposition properties of virtual machine monitor. System states and events from the VMM are intercepted by a policy engine to take a decision of pausing the VM state or refusing access to the hardware resources. The policy engine acts as IDS (intrusion detection system) with strong isolation and also has good visibility into the state of the host that needs to be monitored. Paladin [75, 76] detects the kernel-level rootkit by monitoring the write access to the memory image of the kernel, various jump tables, and system files. StackSafe [77] also checks for the write attempt to the kernel code. OSck [78] detects static control-flow modifying kernel-level rootkits by write protecting kernel text, read-only data and special machine registers. Zhang et al. [79] use Kernel-based virtual machine (KVM) to protect the static kernel code and static kernel data structures against write attempts to those sections.

#### *3.4.1.2 Hashing Known Memory Region*

Rootkit signatures or low-level filesystem scans can be easily fooled by advanced kernel-level rootkit. Unauthorized kernel modification caused by kernel-level rootkit can be detected by checking the periodic hashes of the static data structures and kernel code segment. Pioneer [80] uses a software-based code attestation approach to periodically verify the kernel code segment hashes by SHA-1 hash function. XenKIMONO [81] uses MD5 hashing algorithm to monitor the integrity of kernel text and jump tables. Psyco-Virt [82] computes hashes of critical kernel text using SHA512. RootkitDet [83] registers the kernel and the potential LKMs of the guest OS earlier and performs a comparison of SHA-1 checksums to detect malicious modification of legitimate code by kernel-level rootkits. Patagonix [84] verifies the integrity of all executing binaries by inspecting the code as it executes in the memory using an external database [85]. Another corresponding literature is Kvm-SMA [86]. Kvm-SMA is a security management architecture that monitors the integrity of guest VMs and does not any modification to guest VM. Win et al. [87] proposed to hash only 8 bytes from the initial starting offset of the $9^{th}$ byte to reduce the overhead. EPA-RIMM [88] leverages System Management Mode (SMM), a privileged x86 CPU mode, to measure kernel integrity by periodically checking SHA-256 hash values of particular memory region, control registers and model-specific registers. SGX-Mon [89] leverages Intel's SGX [90] to enclave integrity monitor inside user-space and uses CRC-32, SHA-256 hashing algorithm for performing checksum operation. System call addresses and system call hash values are used in CloudMon [91] to detect kernel-level rootkit in cloud environment. State-based control flow integrity, SBCFI [30] also uses hash function to validate the kernel text including static control flow transfer.

*3.4.1.3   Access Control Policy*

The integrity of the kernel can be protected by imposing access control policy to sensitive kernel objects like kernel text, system call table, interrupt descriptor table etc. The policy module can be easily implemented in VMM layer as it has the higher privilege than the OS kernel. Xu et al. [92] described a flexible and fine-grained access control policy based on the usage control model (UCON) with decision continuity and attribute mutability properties for kernel integrity protection.

*3.4.1.4   Page-level Dynamic Tracing*

A secure system call always executes unmodified pages and modified pages or new allocated pages are executed by a hooked system call. Page-level execution sequence of the system call and the content of these pages are monitored to create a secure control-flow database. Zhan et al. [93] presented a dynamic page-level kernel control-flow integrity checking solution in the cloud.

### 3.4.2  Dynamic Region Integrity

*3.4.2.1   Function Pointers Verification*

Kernel-level rootkit can modify the OS control-flow by using function pointer to point to a malicious code to execute. Kernel-level rootkit can be detected by checking the function pointers if they are pointing to any untrusted code. KOP [48] performs a systematic analysis of function pointers in kernel memory snapshot that can be used to detect kernel-level rootkit. In kernel memory, the EIP register stores the address of the next instruction to be executed and EBP register contains the address located just behind the return address. If the function pointers executed in kernel mode point to an address outside of valid kernel code regions, a kernel control-flow integrity violation is triggered. This approach is used in StackSafe [77] to verify the control-flow integrity. OSck [78] verifies function pointers with the type-graph specified by the kernel code to detect kernel-level rootkit modifying dynamic control-flow. MAS [66] uses memory traversing to verify function pointers pointing to the trusted code. SBCFI [30] considers the dynamic state of the kernel and verifies that function pointers point valid code to validate the dynamic control flow transfer.

*3.4.2.2   Kernel Data Layout Partitioning*

Kernel memory can be partitioned with different access control policy to restrict access to the data in a protected region. Loaded modules and drivers can be restricted to write only driver data and portions of the core kernel data. Only trusted core kernel code is allowed to write any kernel data. In Linux kernel memory the code spans from *_text* to *_etext*. Sentry [94, 95] specifies what data objects can be written in what kernel code regions using kernel memory access control policy.

*3.4.2.3   Secure Page Mapping*

The data that need to be protected are listed in a page table and virtual addresses that have privileges to modify protected dynamic data legally get whitelisted to detect kernel-level rootkit. Any virtual address outside of the whitelist trying to modify protected dynamic data indicates a suspicious attempt by kernel-level rootkit. An instruction trying to modify protected virtual address not registered in the whitelist is skipped. MOSKG [96] implements secure page mapping in multiple OS to protect critical kernel data.

*3.4.2.4 Event-based Behavior Pattern*

Traditional kernel-level rootkits can be analyzed to characterize the malicious behavior patterns of OS events including register accesses, memory accesses, system calls, etc. If any pattern is matched during normal OS runtime, an integrity checker runs to check kernel invariants violation. The static memory region is checked with hash values and the dynamic kernel data are checked with sequences of basic events like in BehaviorKI [97].

Table 5: Integrity-based Detection of Kernel-level Rootkit: Strengths and Challenges/Weaknesses.

| Approaches | Strength | Challenges/Weaknesses |
|---|---|---|
| **Static Region Integrity** | | |
| Write attempt to read-only memory section. | Kernel-level rootkit can be prevented. | Only static region can be protected. |
| Hashing known memory region. | Difficult to fool or tamper the value. | Need to store a hash table. |
| Access control policy. | Integrity of the kernel can be protected. | Policy modules need to be implemented. |
| Page level dynamic tracing. | Improved execution time than branch or instruction level monitoring. | DKOM attack cannot be detected. |
| **Dynamic Region Integrity** | | |
| Function pointer verification. | Static and dynamic function pointers can be verified. | May require OS kernel source code. |
| Kernel data layout partitioning. | Sensitive members of important data structures can be protected. | Requires code revision of OS kernel source code. |
| Secure page mapping. | Can be implemented in different OS. | Whitelist can suffer lack of completeness and the extent of protection is not sufficient. |
| Event-based behavior pattern. | Behavior pattern will trigger the integrity checking. | Event interception will cause performance overhead. |

## 3.5 External Hardware-based Detection

Kernel-level rootkit can also be detected using external hardware devices and the detection system is isolated from the monitored system. Though this approach is not much popular, still there are some effective solutions to detect kernel-level rootkit. This detection approach can be divided into two sub-categories: Snap-based and Snoop-based. Figure 3 shows a simplified overview of external hardware-based detection approach using PCI card. The strength and weaknesses or challenges of the external hardware-based kernel-level rootkit detection approaches are shown in table 6.

*3.5.1 Snap-based Detection.*

*3.5.1.1 Hashing Known Memory Region*

By utilizing a Peripheral Component Interconnect (PCI) add-in card, host memory can be retrieved for examination without the knowledge about or intervention of the host kernel. A monitor is placed inside the add-in card that creates known good hashes for kernel text, text of LKM, and critical data structures and then periodically checks for changes. Copilot [98] is one of the first external hardware-based kernel-level rootkit detection systems. Copilot uses MD5 hashing algorithm and depends on some specific features of the IBM PCI bus. Wang and Dasgupta [99] proposed a kernel-level rootkit detection system that checks part of the OS kernel integrity by external hardware, and which results in checking other static parts of the kernel using cryptographic hash. GRIM [100] leverages GPU architecture to improve the detection rate of snap-based system and shows the impact of multiple hashing algorithm to detection rate.

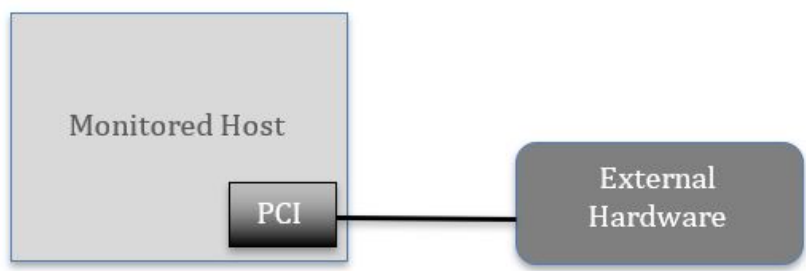

Figure 3: A simplified overview of external hardware-based detection using PCI card.

#### 3.5.1.2 Data Structure Invariants

Sophisticated kernel-level rootkits evolve to tamper kernel dynamic data structures instead of static kernel memory region. An external PCI-based monitor can be used to access low-level kernel data structures of the host and model a set of constraints that will remain correct at runtime for an unmodified kernel. Petroni et al. [101] demonstrated such constraints for detecting kernel-level rootkits. Gibralter [102, 103] also uses external PCI card to hypothesize and infer invariants on kernel data structures to detect kernel-level rootkit.

### 3.5.2 Snoop-based Detection

#### 3.5.2.1 Write Operation to Immutable Region

The operation of the host system can be monitored from an independent system outside the host system by snooping the bus traffic of the host system. Any modification to kernel immutable region of the host OS becomes detectable by snooping the write operation on those addresses. Vigilare [104, 105] is claimed to be the first external hardware-based kernel-level rootkit detection system that has the snooping capability to monitor the kernel integrity.

#### 3.5.2.2 Event Triggered Mutable Object Monitoring

KI-Mon [106] is an event-triggered external hardware-based kernel integrity monitor for mutable kernel objects. To report the address and value pair of memory modification on a monitored object, KI-Mon generates an event. The system detects VFS modification by hardware-assisted whitelisting-based verification events and uses callback-based semantic verification events to detect LKM hiding modification. The authors extended their work [107] on ARM architecture to demonstrate the efficacy in terms of KI-Mon's performance overhead and processor usage.

Table 6: External Hardware-based Detection of Kernel-level Rootkit: Strengths and Challenges/Weaknesses.

| Approaches | Strength | Challenges/Weaknesses |
|---|---|---|
| **Snap-based Detection** | | |
| Hashing known memory region. | Difficult to fool or tamper the value. | Transient attacks can evade detection. |
| Data structure invariants. | Both control and non-control modification can be detected. | OS kernel source code may require, and invariants can be incomplete. |
| **Snoop-based Detection** | | |
| Write operation to immutable region. | Transient attacks can be detected. | Cannot detect DKOM attack. |
| Event-triggered mutable object monitoring. | DKOM attack can be detected. | Additional cost for external hardware. |

## 3.6 Learning-based Detection

With the increase of cybercrime in recent years, the automatic detection of known and unknown attacks now become important in modern security systems. A learning-based detection is an excellent approach to automatically detect known and unknown attacks with high accuracy. Figure 4 shows a general overview of learning-based detection approach. The strength and weaknesses or challenges of the learning-based kernel-level rootkit detection approaches are shown in Table 7. Table 8 shows the summary of the learning-based detection approaches for the kernel-level rootkit.

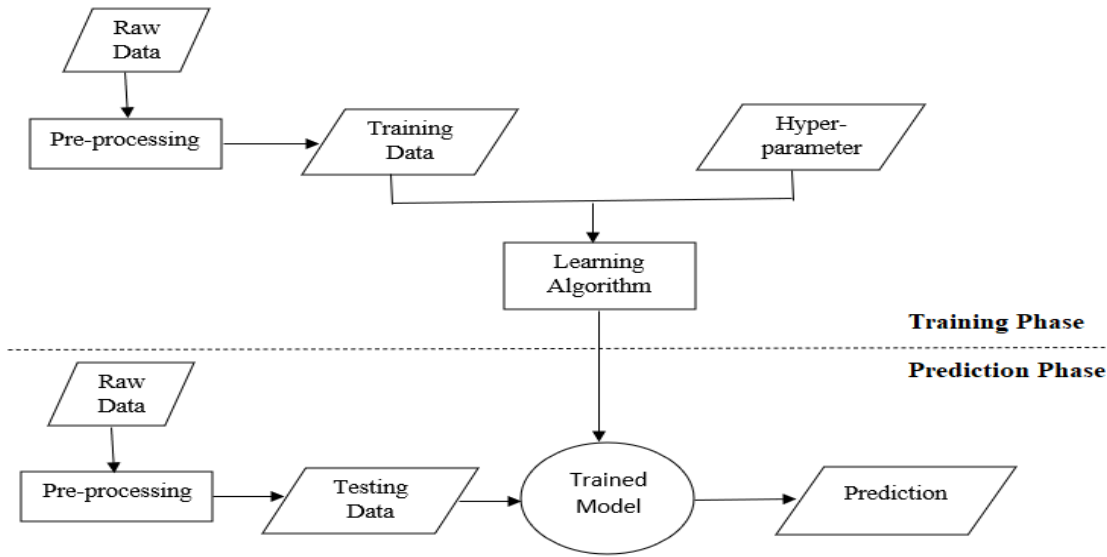

Figure 4: A general overview of learning-based detection approach. In the training phase, learning models is trained using training data and optimized using hyper-parameters. The trained model is then used to predict the output of new data fed into the system.

### 3.6.1 Emulating Kernel Driver Behavior

Learning algorithm can be applied to a set of kernel driver run-time features derived from the execution behavior using emulator to distinguish between malicious and legitimate kernel drivers. Limbo [108] is more likely a preventive approach that forces the kernel driver to execute in an emulated environment and extract the features of the kernel driver. Selection of kernel driver features is based on their run-time behaviors and binary attributes. Limbo used a Naïve Bayes classifier training tool to distinguish between legitimate and malicious Windows kernel drivers with the extracted features as input. The author classified the features into seven categories of which each member's value is either a logical value (true or false) or an integer count. As the Limbo executes the kernel driver in the emulator to extract features, it poses additional delay in loading time of the kernel driver.

### 3.6.2 Statically Analyzing Kernel Driver

The obfuscation employed in kernel-level rootkit binaries makes the static analysis difficult. Still kernel-level rootkit can be detected through static analysis by disassembling the kernel driver and extract features like general behavior, communications, suspicious behaviors etc. Musavi and Kharrazi [109] focused on static analysis to detect kernel-level rootkit. When a user-level application installs or drops a driver, the detection process disassembles the driver to extract a set of features and use a binary classifier to distinguish between malicious and legitimate drivers.

### 3.6.3 Virtual Memory Access Pattern

Memory access pattern of legitimate and infected execution of an application differ if kernel-level rootkit modifies associate control-flow or data structures. Instead of distinguishing malicious and benign applications, Xu et al. [110] proposed to use virtual memory access pattern to distinguish exploited execution and legitimate execution of each application. For each system call, four types of memory accesses are used as feature set to train the machine learning model.

### 3.6.4 Event Counts Using Hardware Performance Counter

Events associated with hardware related activities such as clock cycles, cache hits/misses, branch behavior, memory resource access patterns etc. can be counted using HPC. The events count will differ from normal counts if kernel-level rootkit modifies the control-flow of the OS kernel. This approach will not work against DKOM attack as no malicious code will be executed during trace-collection. Singh et al. [111] designed five different synthetic rootkits with single rootkit functionality and used those rootkits to identify the most important HPCs. The authors used four machine learning classifiers (SVM, OC-SVM, Naïve Bayes, and Decision Tree) to train the machine learning model with HPC traces data.

### 3.6.5 Volatile Memory Traces

Memory forensic analysis can also be combined with learning-based approach to detect kernel-level rootkit. Volatility [59] plugins can be used to extract features from memory dumps. The extracted features may include hidden kernel modules, abnormal driver objects, SSDT hooking, abnormal callbacks and timers, orphan threads, and other hooking behaviors. TKRD [112] experimented with memory dump features using seven machine learning classifiers and evaluated their performance. Nadim et al. [139, 140] also proposed characteristic features of the kernel-level rootkit extracted from volatile memory traces to train learning-based models.

### 3.6.6 Access Operation to Code, Data, and Register

The run time behavior of a kernel module can be divided into three following categories: code access, data access, and hardware register access. Hardware assisted virtualization technique can be used to isolate memory region and registers access for a kernel module, and then the behavior of that kernel module can be extracted. The behavior features of kernel module may include important kernel API invocation, executing code in kernel data region, write operation to kernel memory area, write operation to important hardware registers etc. VKRD [113] experimented with these features to train multiple machine learning algorithms. As the features are either binary or a counter value, they used Min-Max normalization method to normalize the values.

### 3.6.7 System Call Execution Time

Since a large number of kernel-level rootkits modify the control flow by altering system calls, system call times can be an important feature to detect kernel-level rootkit. Luckett et al. [114] proposed a behavior-based analysis of system call execution times. The authors used the neural networks to classify system calls for detecting the presence of rootkit within a system.

### 3.6.8 Process Execution Behavior Profile

Deviation from execution behavior profiles of dynamic intra-process based on architecture level semantics can be used to detect kernel-level rootkits. The key insight of this mechanism is that the kernel-level rootkit leaves abnormal traces in architecture-level semantics by maliciously modifying the kernel objects that distort the execution flow of benign processes. Hardware events like data dependencies between registers, OS privilege transition, and branches in program execution flow can be incorporated to interpret the program data/control transfer flow as features. Zhou and Makris [115] introduced a hardware-assisted machine learning-based rootkit detection mechanism that first identifies the process class and then employs Kernel Density Estimation (KDE) to indicate a compromise in process behavior caused by a kernel-level rootkit.

Table 7: Learning-based Detection of Kernel-level Rootkit: Strengths and Challenges/Weaknesses.

| Approaches | Strength | Challenges/Weaknesses |
| --- | --- | --- |
| Emulating kernel driver behavior. | Prevent malicious driver to load. | Additional delay in driver loading time. |
| Statically analyzing kernel driver. | Analysis can be done inside the host. | Detector is vulnerable to advanced kernel-level rootkit. |
| Virtual memory access pattern. | Malware leaves fingerprints on program memory accesses. | DKOM attacks may remain undetected. |
| Events count using HPC. | Control-flow modification can be detected with high accuracy. | DKOM attacks have no impact on HPC. |
| Volatile memory traces. | Detection system can be implemented separately. | Transient attacks can evade detection. |
| Access operation to code, data, and registers. | Target kernel module can be isolated from kernel space. | Memory isolation may introduce significant performance overhead. |
| System call execution times. | System calls need to be executed to perform malicious activities. | May have no impact on DKOM attack. |
| Process execution behavior profile. | Immune to software tampering. | Hardware assistance will cause performance overhead. |

Table 8: Summary of Learning-based Kernel-level Rootkit Detection Approaches.

| Prior work | Feature | Learning Algorithm | Operating System |
|---|---|---|---|
| Limbo [108] | Static attributes of driver's binary and dynamic attributes like data structure access, descriptor table, and driver-related features. | Naïve Bayes | Windows |
| Musavi and Kharrazi [109] | Dis-assembled driver's code including Kernel function calls, constants, assembly commands, variable type etc. | C5 Tree | Windows |
| Xu et al. [110] | Virtual memory access pattern of system call. | Random Forest, SVM, Logistic Regression | Linux Debian |
| Singh et al. [111] | Event count using Hardware Performance Counter (HPC). | SVM, OC-SVM, Naïve Bayes, Decision Tree | Windows 7 |
| TKRD [112] | Volatile memory traces of modules, threads, drivers, IRP and SSDT hooks, callbacks, and timers. | Random Forest, J84, JRip, PART, BayesNet, Naïve Bayes, SMO | Windows 7 |
| VKRD [113] | Run-time features of kernel modules such as Kernel API invocation, Code write, Data write, and Register access operations. | SVM, Decision Tree, Random Forest, KNN | Windows XP |
| Luckett et al. [114] | System call execution time. | Feed Forward, Nonlinear auto-regressive | Linux Ubuntu |
| Zhou and Makris [115] | Data dependencies on general purpose registers and branches in program execution flow. | KNN, SVM, ANN | Linux |

## 4 MORE KERNEL-LEVEL ROOTKIT LITERATURES

In this section we will discuss about the prior literature on preventing kernel-level rootkit and profiling the kernel-level rootkit behavior and widely used tools for detecting kernel-level rootkit.

### 4.1 Kernel-level Rootkit Prevention

Zhao et al. [81] proposed a secure virtual file system (SVFS), a prevention system that provides secure data storage against a kernel-level rootkit. SVFS stores sensitive files in a dedicated virtual machine separate from application guest virtual machines. All the accesses to sensitive data are subject to be applied by access control policy when going through SVFS. Therefore, the kernel-level rootkits cannot bypass this protection by compromising application guest OS. The limitation of SVFS is that it does not prevent kernel-level rootkit to exploit guest OS, it only prevents kernel-level rootkits to run automatically when guest OS reboots.

Seshadri et al. [116] formulated SecVisor to ensure code integrity for OS kernels by allowing only user-approved code to execute in kernel mode. Hardware memory protections are used to ensure kernel code integrity. Both CPU's memory management unit (MMU) and I/O memory management unit (IOMMU) are modified to ensure that only kernel code confirmed by a user-supplied policy will be executed. By these modifications, the kernel can be protected against malicious writes via direct memory access (DMA) device. SecVisor works as a preventive tool against kernel-level rootkit after loading themselves into the memory. However, if the OS kernel has pages that contain both data and code, SecVisor does not function. Additionally, SecVisor requires modifying the source code of the kernel, which makes it difficult to support for closed source operating systems like Windows.

Butler et al. [117] introduced a rootkit-resistant disk (RRD) that label all configuration files and system binaries to prevent a compromised operating system from infecting its on-disk image. The RRD is implemented on a network storage device not to make the kernel-level rootkit become persistent. A tightly governed administrative token required for system write-capability blocks any malicious modification of the immutable memory block of the host OS during normal operation.

NICKLE is a virtual machine monitor (VMM) based kernel-level rootkit detection and prevention system presented by Riley et al [118]. It uses a memory shadowing scheme to store the authenticated kernel code in the shadow memory and at the runtime, transparently routes guest kernel instruction fetches to the shadow memory. The NICKLE system effectively works in Linux and Windows OSes targeting kernel-level rootkit. As NICKLE does not modify kernel code, it easily overcomes the drawbacks of SecVisor. However, NICKLE does not effectively protect the self-modifying kernel code, which is available in both Linux and Windows OS and does not support kernel page swapping.

One of the most commonly adopted techniques by kernel-level rootkits to evade detection is hooking the kernel object of the system. To efficiently protect the kernel hooks from being hijacked in a guest OS, Wang et al. [119] proposed HookSafe that relocates kernel hooks to a dedicated page-aligned memory space. Then the accesses to the kernel hooks are regulated with hardware-based page-level protection. Besides memory-based kernel hooks, HookSafe also regulates the accesses of hardware registers such as Interrupt Descriptor Table Register (IDTR), Global Descriptor Table Register (GDTR), SYSENTER MSR registers, and DR0-DR7 debug registers. The system successfully prevents modification of protected kernel hooks against real-world kernel-level rootkits.

Oliveira and Wu [120] proposed a solution that protects kernel code and data integrity by preventing kernel-level rootkits. At the architecture level (memory and registers), all the write attempts to kernel code and data segments are checked for validity by enforcing Biba's star [121]. The process associated with the illegal write operation is terminated but the rest of the system is allowed to continue execution.

Xuan et al. [122] presented DARK, a system that tracks LKM to prevent kernel-level rootkits. By dynamically switching a running system between virtualized and emulated execution, DARK thoroughly captures the target module's activity in a guest OS. It provides a flexible security policy framework with access control rules to detect malicious modules. The kernel rules are then experimented against kernel-level rootkits to find out effectiveness.

Rootkits often reside in the storage to survive from system reboots thus, pose a serious security threat being persistent. A hypervisor-based file protection scheme was presented by Chubachi et al. [123] to prevent persistent rootkits from residing in the storage. The authors run the target OS without hypervisor to create a security policy and map protected files to a set of regions in the storage with administrator mode. By making the critical file system always read-only, the target OS is then run with a hypervisor in normal mode. As the hypervisor has a higher privilege than the target OS's kernel, kernel-level rootkits are not able to overwrite the security policy by manipulating the kernel.

Grace et al. [124] introduced a hardware virtualization-based architecture to protect commodity OS kernel against kernel-level rootkits. This prevention system can effectively reduce performance overhead without modifying the commodity OS kernel. The authors use page-level redirection of instruction fetches and make them mode-sensitive by redirecting only kernel instruction fetches. However, the proposed prevention system does not protect kernel control-flow integrity and does not support self-modifying kernel code.

Schmidt et al. [125] presented an approach to prevent kernel-level rootkit attacks as well as to detect malware in the cloud computing environment. To load only cryptographically authorized and trusted kernel modules, the OS kernel is modified. By checking the integrity of the authorized kernel modules, kernel-level rootkit attacks through malicious modules can be prevented.

### 4.2 Profiling Kernel-level Rootkit Behavior

To design an effective kernel-level rootkit detection solution, it is important to profile best behaviors that reveal kernel-level rootkits. The system proposed by Levine et al. [20] not only detects the kernel-level rootkits but also categorizes detected kernel-level rootkits based on the assumption that for a particular kernel-level rootkit, the implementation of each malicious system call is uniform. From the archived hash values of malicious system calls, they categorize a new unknown kernel-level rootkit to a modified version of previously known kernel-level rootkit or a new one. They conclude that a new kernel-level rootkit retrieved from honeynet is a combination of two previously known rootkits [126].

One of important kernel-level rootkit's tasks is to execute malicious code that manipulates the sensitive data accessed by user-level programs to reflect system states via system calls or critical data structures maintained by the kernel. K-Tracer, proposed by Lanzi et al. [127], is a dynamic kernel-level analysis engine for the Windows OS that performs data-flow analysis on sensitive data to extract the malicious behavior of kernel-level rootkit. To identify the rootkit behavior, K-Tracer uses a combination of forward and backward slicing techniques on selective stimulated kernel events. K-Tracer was implemented on the QEMU [128] emulator environment to perform instruction-level execution tracing, leaving a probability of evasion by malware that can detect underlying emulator [129]. This approach also has some limitations against sophisticated rootkit techniques such as DKOM (direct kernel object modification) for which authors discussed further improvement of the system to counter such sophisticated kernel-level rootkits.

Wang et al. [130] proposed a systematic approach named HookMap to identify the kernel hooks used for hiding the presence of rootkits. By their design, kernel-level rootkits attempt to conceal their presence from various system utility programs. HookMap analyzes the kernel side execution path of those programs to find the set of kernel hook that are potentially vulnerable for attack by kernel-level rootkits. The authors manually analyzed Linux-based rootkits and found that all identified kernel hooks are listed in their results. This approach is only effective when applying to the kernel-level rootkits that attack the kernel control flow.

HookFinder, a prototype developed by Yin et al. [131] automatically identifies hooking behavior of malicious code and extract hook implementation mechanisms without any prior knowledge. To identify a hook, they observe the instruction pointer. The change in memory with other machine states are labeled as *impact*. If the instruction pointer is loaded with marked *impact* and the execution jumps immediately into the malicious code, they identify the hook. An emulator is used for implementing the HookFinder, which provides isolation between the analysis environment and the malware.

PoKeR, a virtualization-based kernel-level rootkit profiler introduced by Riley et al. [132] is comprised of four aspects: hooking behavior, targeted kernel object, user-level impact, and injected code. It profiles not only traditional system call hook-based rootkits but also DKOM-based rootkits. To accurately determine the kernel objects that are modified by a kernel-level rootkit, PoKeR uses a combat tracking technique that maintains a map of dynamic kernel objects. The authors used NICKLE as the detection system to generate a kernel-level rootkit detection point.

Rkprofiler [133], an analysis and profiling system for Windows OS kernel running in a VM, inspects each instruction executed and captures all function calls to construct a call graph for kernel malware execution. It also tracks dynamic data objects and hardware access events of kernel malware. With the extracted information, Rkprofiler reports the kernel malware behavior in a guest OS. DORF, Data Only Rootkit Framework [134] is an object-oriented framework designed by Ryan Riley that allows researchers to prototype and test data only kernel-level rootkit attacks in various Linux distributions and versions. The author also divided the kernel-level rootkit attacks based on their influence and clarified their definitions to defend them. Using the DORF prototype, researchers can easily test their developed defense system against various kernel-level rootkits. Kernel-level rootkits not only modify user-level activities like system call and APIs but also modify kernel-level activities. MrKIP, a system developed by Wang et at. [135], semi-automatically profiles kernel-space activities of kernel-level rootkits. The invocations of important in-kernel functions with the associated arguments construct the behavior profile. New variants of rootkit families can be recognized with those collected behavior profiles.

HProve [136] is a hypervisor level provenance tracing system that reveals causality dependencies among kernel-level rootkit behaviors and impacts on the victim system by replaying the kernel-level rootkit attack. The proposed system records the whole system execution of the guest OS through a lightweight manner and keeps track of a series of kernel functions and memory access traces to sensitive kernel objects.

## 5 FUTURE RESEARCH DIRECTIONS

Many approaches have been proposed including the learning-based approaches to successfully detect and prevent the kernel-level rootkit. However, still many challenges need to be addressed that are crucial for the high accuracy of kernel-level rootkit detection. In this section, we present conceivable forthcoming research directions that can be considered by the researchers as a future work.

- *A. Artificial Intelligence*: Artificial intelligence (AI) methods have shown their success in countless domains to learn complex systems and make an informed decision. This is an umbrella term under which machine learning and deep learning take place. Though there are few published research in the kernel-level rootkit detection domain using AI, it is still not the most popular approach in this domain. Most of the published works in this domain either suffer to detect the DKOM attacks or introduce performance overhead. Overcoming these drawbacks can be a direction to future research. Unfortunately, there has been a lack of open-source dataset for kernel-level rootkit detection. The prior work of the kernel-level rootkit detection in AI used their own dataset, which are not available for others. A standardized and updated publicly available dataset is required to perform detection analysis in an efficient way. Future research will look into building an open-source dataset for kernel-level rootkit detection resulting in detecting unknown new attacks by training an AI model. Additionally, because the characteristic features of the kernel-level rootkits are continuously evolving, the training data set should dynamically include new samples using incremental learning to make the AI model remain effective.

B. *Container Environment:* In recent years, container-based service has been increasingly deployed by the service providers for its flexibility and efficiency. We can define a container as a software unit with all dependencies installed that helps applications to run quickly and reliably [40]. Unlike the virtual machines, containers are isolated using kernel functionalities such as namespace, c-group, etc. Despite its benefit of the portability and the ease of deployment, the container is less secure than the fully isolated virtual machines. The isolation of the container can be invalidated when the kernel-level rootkits exploit vulnerabilities existing in the kernel. This may lead to critical security incidents that need to be addressed as a future work.

C. *Zero-Day Attack Detect*: Most of the current approaches of detecting the kernel-level rootkit are postmortem type. They only detect the kernel-level rootkit after the intruders compromise the system. Because it is quite difficult to predict the attack scenario, a highly intelligent and lightweight approach is required to examine the OS behavior at run time and detect a zero-day attack.

## 6 CONCLUSION

A systematic literature survey of the kernel-level rootkit detection approaches is presented in this paper. The reviewed papers have been cautiously investigated to provide a broad and structured solution taxonomy for the kernel-level rootkit detection. The detection approach of the kernel-level rootkit is classified into six main categories: Signature-based, Behavior-based, Cross-view-based, Integrity-based, External hardware-based, and Learning-based. The strengths and weaknesses or challenges of each detection approach are identified in this paper. Most of the prior kernel-level rootkit detection approaches are cross-view-based and integrity-based. Learning-based detection has been proposed in the last few years. This detection is sub-categorized based on the features used to train the learning model. The prevention techniques against the kernel-level rootkit in prior literatures are also reviewed along with the literatures about profiling of the kernel-level rootkit behavior. This work introduced a broad overview of the kernel-level rootkit detection, prevention, and behavior profiling for the future research.